\documentclass{ws-procs975x65}

\begin{document}

\title{SAGACE: \\ the Spectroscopic Active Galaxies And Clusters Explorer}

\author{P. DE BERNARDIS $^{1*}$    ,
D. BAGLIANI$^2$, A. BARDI$^3$, E. BATTISTELLI$^1$, M.
BIRKINSHAW$^4$, M. CALVO$^1$, S. COLAFRANCESCO$^5$, A. CONTE$^1$,
S. DE GREGORI$^1$, M. DE PETRIS$^1$, G. DE ZOTTI$^6$, A.
DONATI$^3$, L. FERRARI$^2$, A. FRANCESCHINI$^6$, F. GATTI$^2$, M.
GERVASI$^7$, P. GIOMMI$^5$, C. GIORDANO$^1$, J.
GONZALEZ-NUEVO$^8$, L. LAMAGNA$^1$, A. LAPI$^9$, G. LUZZI$^1$, R.
MAIOLINO$^{10}$, P. MARCHEGIANI$^5$, A. MARIANI$^3$, S. MASI$^1$,
M. MASSARDI$^8$, P. MAUSKOPF$^{11}$, F. NATI$^1$, L. NATI$^1$, P.
NATOLI$^{9}$, M. NEGRELLO$^{12}$,  F. PIACENTINI$^1$, G.
POLENTA$^1$, M. SALATINO$^1$, G. SAVINI$^{13}$, A. SCHILLACI$^1$,
S. SPINELLI$^7$, A. TARTARI$^7$, M. TAVANTI$^3$, A. TORTORA$^3$,
M. VACCARI$^6$, R. VACCARONE$^2$, M. ZANNONI$^7$, V. ZOLESI$^3$}

\address{
$^1$Dipartimento di Fisica, Universit\'a La Sapienza, and INFN
sezione di Roma, Roma, Italy; $^2$Dipartimento di Fisica,
Universit\'a di Genova , and INFN sezione di Genova, Italy;
$^3$Kayser Italia, Livorno, Italy; $^4$Department of Physics,
University of Bristol, UK; $^5$ASDC - ASI - Frascati , Italy, $^6$
INAF, Osservatorio di Padova, Italy; $^7$Dipartimento di Fisica,
Universit\'a di Milano Bicocca, Italy; $^8$SISSA - Trieste, Italy;
$^{9}$Dipartimento di Fisica, Universit\'a di Tor Vergata, Roma,
Italy; $^{10}$INAF - Osservatorio di Roma, Italy;
$^{11}$Department of Physics and Astronomy, Cardiff University,
UK; $^{12}$Department of Physics and Astronomy, Open University,
Milton Keynes, UK; $^{13}$Department of Physics and Astronomy,
University College London, UK; $^{*}$e-mail:
paolo.debernardis@roma1.infn.it}

\begin{abstract}
The SAGACE experiment consists of a mm/sub-mm telescope with a 3-m
diameter primary mirror, coupled to a cryogenic multi-beam
differential spectrometer. SAGACE explores the sky in the 100-760
GHz frequency range, using four diffraction-limited bolometer
arrays. The instrument is designed to perform spectroscopic
surveys of the Sunyaev-Zeldovich effects of thousands of galaxy
clusters, of the spectral energy distribution of active galactic
nuclei, and of the [CII] line of a thousand galaxies in the
redshift desert. In 2008 a full phase-A study for a national small
mission was completed and delivered to the Italian Space Agency
(ASI). We have shown that taking advantage of the differential
operation of the Fourier Transform Spectrometer, this ambitious
instrument can operate from a Molniya orbit, and can be built and
operated within the tight budget of a small mission.

\end{abstract}


\bodymatter

\section{SAGACE Science}\label{aba:sec1}

Our knowledge of the mm/sub-mm sky is rapidly improving.
Balloon-borne missions and the recent WMAP all-sky survey have
produced impressive maps of the cosmic microwave background (CMB),
allowing the precise measurement of several cosmological
parameters. The Planck \cite{Lama03, Lawr03} mission is producing
maps of the whole sky in nine wavebands with exquisite accuracy,
allowing a secure separation of the CMB signals from the Galactic
and extragalactic foregrounds, and a massive shallow survey of
Sunyaev-Zeldovich clusters.

However, the spectroscopic exploration of the frequency range
between 90 and 600~GHz is still completely undeveloped, since most
of this range is not accessible from the ground. In the past, the
FIRAS survey on COBE \cite{Fixs96} has produced coarse
($10^\circ$) spectral maps of the sky with an absolute Fourier
Transform Spectrometer (FTS). The main components in the measured
spectra are the CMB, the continuum emission of diffuse
interstellar dust, and a few prominent lines from the interstellar
medium ([CII], [OIII], ...). We simply do not have
higher-resolution surveys with significant sky coverage, and the
currently-operating Herschel spectroscopic instrument (SPIRE) will
cover only frequencies higher than 450 GHz \cite{Grif09} with a
small field of view.

Diffuse emission in the mm/sub-mm range is very rich in
astrophysical and cosmological information. This motivated our
proposal for a 3-m space telescope coupled to a Fourier Transform
Spectrometer, covering the range 100-760~GHz with four arrays of
diffraction-limited, photon-noise-limited cryogenic bolometers. In
the following we describe a few of the many topics accessible with
a high sensitivity spectroscopic survey of diffuse emission at
these frequencies. Then, we will outline our baseline proposal for
a small (cost-wise) mission, with ambitious goals.

\subsection{Sunyaev-Zeldovich effect}

A fundamental topic of current cosmological research is the study
of the formation and evolution of cosmic structures.
Clusters of galaxies represent an extremely important
structural level in this framework. They are the largest
gravitationally-bound structures in the Universe, and can contain
up to a few thousand galaxies. The cluster volume between galaxies
is filled with a hot (10$^7$-10$^8$~K), ionized gas which makes up a
significant fraction ($\sim 10$-$20$ \%) of the total mass of the
cluster. In addition, the presence of a dominant dark mass component
is required to explain the motions of galaxies in clusters and the
gravitational lensing of background galaxies.

The presence of ionized gas in the intracluster medium is evident
from X-ray observations of clusters of galaxies. Here clusters
appear as diffuse sources, with the ionised gas filling the
potential well of dark matter, heated to millions of degrees, and
producing intense thermal bremsstrahlung emission.  The same
particles of the ionized intracluster gas interact with the CMB
via the inverse-Compton effect, the up-scattering of low-energy
photons off more energetic hot electrons. This phenomenon, also
known as the Sunyaev-Zeldovich Effect (SZE), produces an
additional source of anisotropy in the CMB in the direction of
rich clusters of galaxies.

The optical depth for this effect is small, but not negligible,
because although the density of electrons is only of order
$n_e \sim 10^{-3}$ cm$^{-3}$, the path length $\ell$ through a cluster
medium can be several Mpc.
With a Thomson cross section $\sigma = 6.65 \times 10^{-25}$
cm$^2$, this produces an optical depth $\tau = n_e \sigma \ell
\sim 0.01$. So we have a 1\% probability that a CMB photon
crossing a rich cluster is scattered by an electron. Since the
electron energy is much larger that the energy of the photon, to
first order the fractional energy gain for the photon is about
$kT_e/m_ec^2 = 1\%$. The resulting fractional temperature change
of the CMB is of the order of $1\% \times 1\% = 10^{-4}$,
corresponding to an anisotropy of about 300~$\mu$K in the direction of
the cluster. This is a large signal relative to the tens of $\mu$K
level of the CMB anisotropy, and the few $\mu$K component of this at
arcmin angular scales. This is the thermal part of the SZE.

Being a scattering effect, the SZE does not depend on the distance
of the cluster. For high-redshift clusters, the decrease in solid
angle with increasing redshift is exactly compensated by the
increase of the local temperature of the CMB. Therefore distant
clusters of galaxies that are too faint to be detected in the
X-ray or optical bands can still be observable via the SZE.

\begin{figure}
\center \psfig{file=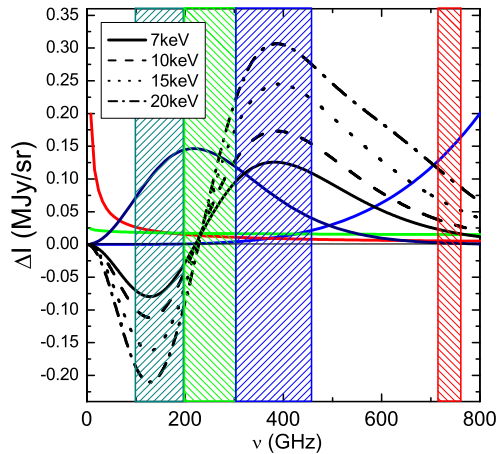,width=3in} \caption{Spectrum of
the thermal SZE in clusters of galaxies.
Plotted (black lines) is the difference between the CMB spectrum
through the cluster and the CMB spectrum outside the cluster. The
different black lines refer to different temperatures of the
intracluster plasma. The other lines represent the spectral shapes
(not normalized) of different contaminants: synchrotron
(red), free-free (green), CMB anisotropy and kinetic SZE (dark
blue), dust (blue). This shows that the thermal SZE spectrum
can easily be distinguished from contaminants by measurements over
sufficient spectral range. The shaded areas
represent the four observation sub-bands of the SAGACE
instrument.} \label{aba:fig1}
\end{figure}

The spectrum of the thermal SZE has a characteristic shape.
Since all interacting CMB photons get approximately a
1\% boost in energy, the result is a transfer of photons in the
CMB spectrum from lower to higher frequencies, resulting
in a decrease of brightness at low frequencies (below 217 GHz) and
an increase of brightness at high frequencies (above 217 GHz), as
shown in Fig.~\ref{aba:fig1}.

Thus the same cluster will be seen as a dark spot in the CMB at
frequencies below 217~GHz, and as a bright spot at frequencies
above 217~GHz. This unusual spectrum makes it possible to extract
the SZE signal even in the presence of contaminating sources, like
confusion from the CMB anisotropy itself, Galactic emission, and
unresolved extragalactic sources.

In addition to the thermal signal, other sources of SZE can be
produced in clusters via the interaction of non-thermal
(relativistic and sub-relativistic) particle distributions with
the CMB photons, providing SZE signals with specific spectral and
spatial characteristics \cite{Cola07}.

The key strategy to perform the SZE signal extraction is to have
spectral coverage of both the negative and positive sides of the
SZE.

A spectroscopic mission like SAGACE is optimized to extract the
SZE from all the other diffuse components, even in distant and
low-mass clusters. Operating from space, SAGACE will not be
affected by atmospheric absorption and noise, the main limiting
factors for current wide-field deep surveys of SZE clusters. The
number of independent bands observable from the ground is limited
in number and spectral range. This results in bias on and
degeneracies between the different cluster parameters recovered
from SZE measurements (mainly the optical depth, the peculiar
velocity, the temperature of the gas). All these problems are
solved by a wide coverage spectrometer (see Sec. \ref{aba:perf}).
Furthermore, the coverage of the whole interesting spectral range
with a single instrument will solve the major problem of the
cross-calibration of different instruments affecting current
measurements.

\subsection{Star forming galaxies at the peak of cosmic activity}

The density of cosmic star formation rate peaks at $z \sim 1.5$.
This is the epoch where galaxies form and assemble at the highest
rate, either through merging or through enhanced gas accretion.
Paradoxically, the redshift interval $1.2<z<2$ is the most
difficult to investigate spectroscopically from the ground.
Indeed, while optical spectrographs with high multiplexing have
been effective in identifying the redshift of large samples of
galaxies at $z<1.2$ and at $z>2$, at intermediate redshifts (the
so called ``redshift desert'') the most prominent emission
features (e.g. Ly-$\alpha$, [OII],...) used to spectroscopically
identify galaxies are outside the optical band. This has hampered
studies of  galaxies in the most important redshift range.
Currently, less than 10\% of star forming galaxies have been
spectroscopically identified within the redshift desert
\cite{Renz09}.

\begin{figure}
\center \psfig{file=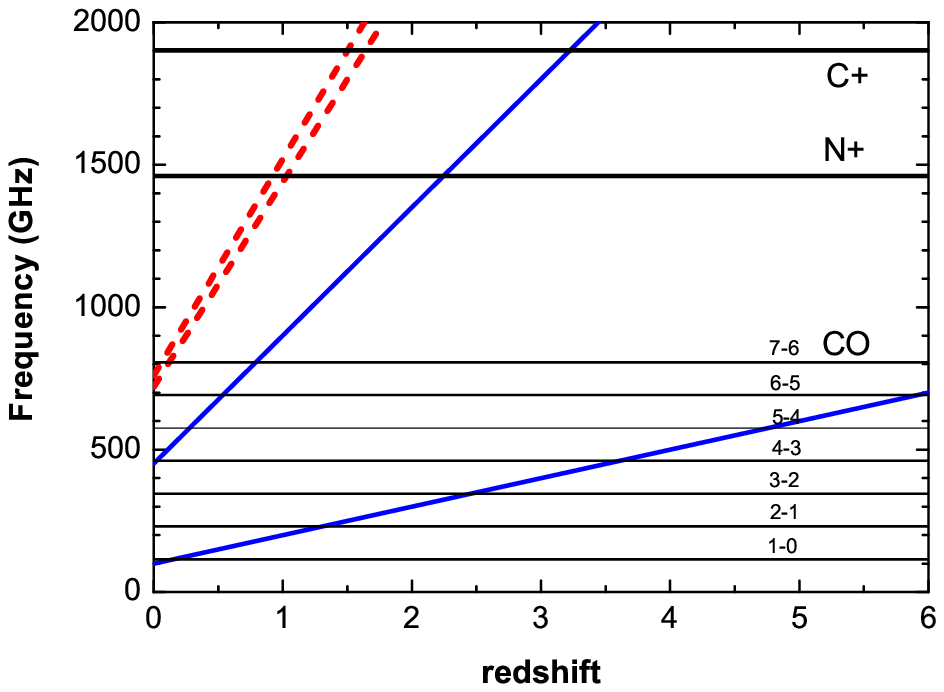,width=2.5in, height=2.0in}
\psfig{file=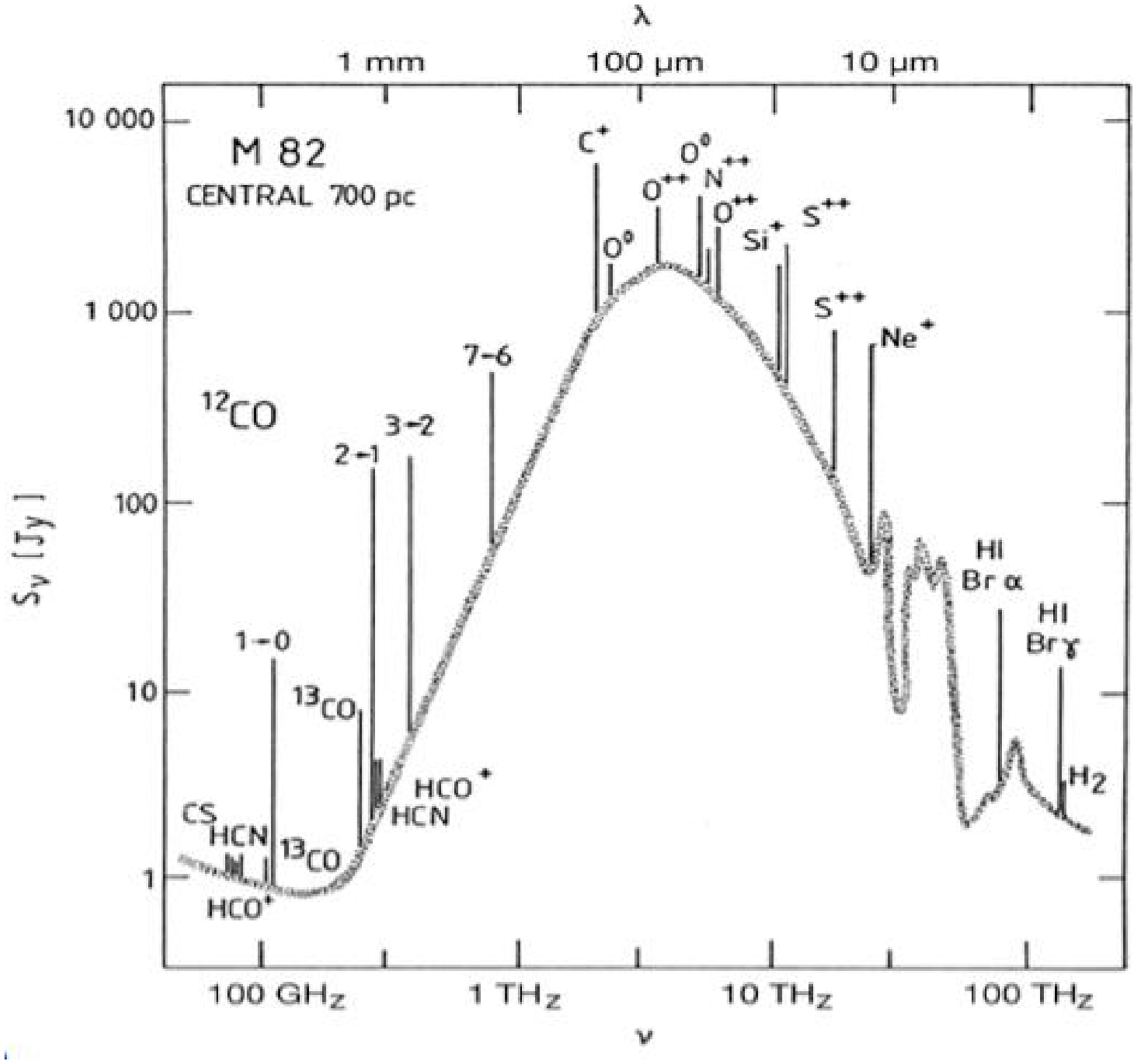,width=2in} \caption{Left: Redshift
dependance of the frequency coverage of SAGACE, compared to the
restframe frequencies of important lines from the diffuse medium
in galaxies. The bands explored by SAGACE lie between the two
solid (blue) lines and between the two dashed (red) lines. Right:
Spectrum of M82, showing the continuum and the lines used in the
left panel. The 158-$\rm \mu m$ C$^+$ line is the brightest
cooling line in galaxies.} \label{aba:fig2}
\end{figure}

The far-IR spectral band hosts the brightest lines in the spectrum
of any galaxy (Fig. \ref{aba:fig2}). In particular the [CII] line
at 158$\ \mu \rm m$ is generally the strongest line in the
spectrum of nearly all galaxies, accounting for as much as 1\% of
the galaxy bolometric luminosity. The use of this and other far-IR
lines is currently limited to the local Universe (through space
missions such as ISO and Herschel \cite{Luhm03}) and to a few very
high redshift targets ($z>4$) for which such far-IR lines are
redshifted into the (sub)mm atmospheric windows
\cite{Maio05,Maio06,Iono06} . The lack of space observatories with
sensitive spectroscopic capabilities in 700-800 GHz has prevented
astronomers from exploiting these emission lines to identify
galaxies around the epoch of peak cosmic star formation.

The high resolution spectroscopic mode of SAGACE at high
frequencies (720-760 GHz) will allow us to identify the redshift
of large samples of star forming galaxies at $1.5<z<1.6$, well
within the redshift desert, by detecting their [CII] line (Fig.
\ref{aba:fig2}). In particular the fast mapping speed of SAGACE at
this frequency will allow the detection of several thousands of
galaxies in this redshift range. The [CII] line will not only
provide the redshifts of the sources but also an indication of
their star formation rate (since [CII] is the main coolant of the
ISM). By detecting thousands of galaxies it will also be possible
to determine the three-dimensional clustering of galaxies at $z
\sim 1.5$, which will allow us to trace the evolution of cosmic
structures around this crucial epoch with unprecedented detail,
with important cosmological implications.

\begin{figure}
\center \psfig{file=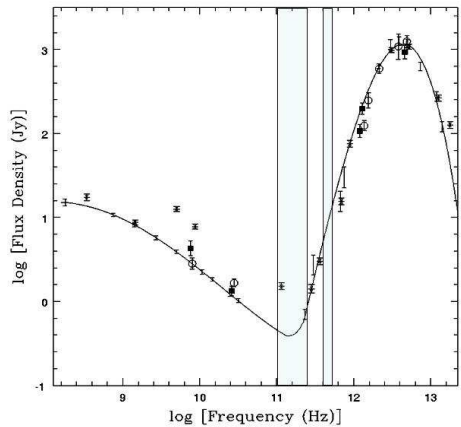,width=2.5in} \caption{ The
continuous line represents the best fit spectral energy
distribution (SED) of the prototypical starburst galaxy
(M82)\cite{Yun00}. The different symbols indicate the measurements
for several quasars and luminous infrared galaxies, normalized to
fit the peak emission of M82. The spectral coverage of SAGACE is
represented by the two shaded areas, and includes the transition
from synchrotron-dominated to dust-dominated emission. }
\label{aba:fig3}
\end{figure}

\subsection{Active Galactic Nuclei}

SAGACE will also perform a wide area ($>$ 1000 deg$^2$) deep
photometric survey down to the confusion limit, thus providing a
unique database for investigating the cosmological evolution of
the luminosity function and of the spectral energy distribution of
Active Galactic Nuclei (AGN) and of star-forming galaxies over a
broad redshift interval, in the poorly explored but crucial
microwave-to-submillimeter wavelength range (see
Fig.~\ref{aba:fig3}).

\section{The SAGACE Instrument}

\label{instrument baseline}

SAGACE has been proposed as a small mission, answering a recent
call of the Italian Space Agency. The design is the result of a
trade-off between scientific ambitions and severe limitations of
weight, complexity and cost.

\begin{figure}
\psfig{file=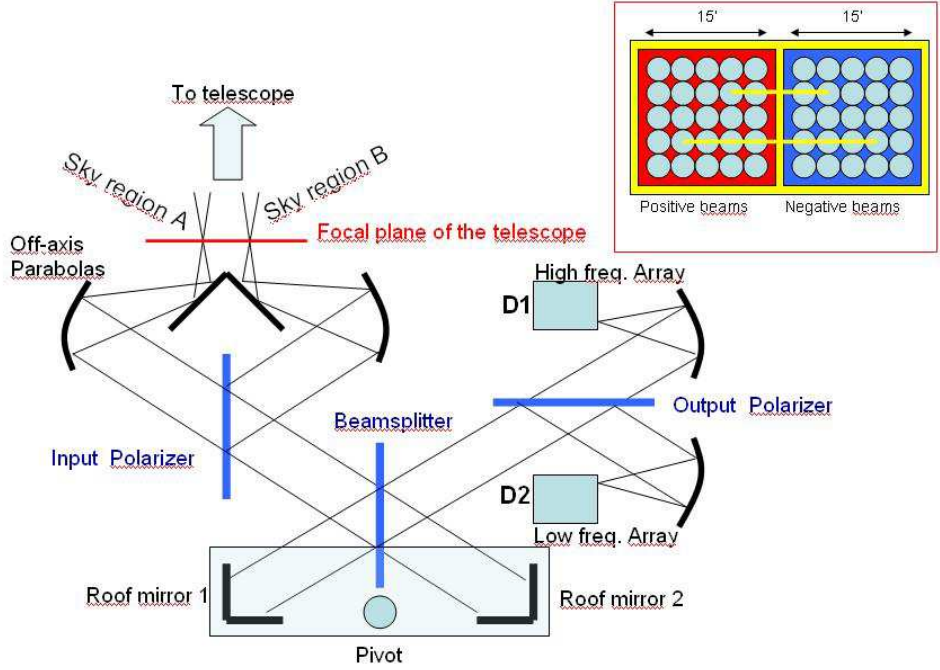,width=5in} \caption{Conceptual diagram of
the SAGACE spectrometer and of its focal plane configuration. The
two inputs of a Martin-Puplett polarizing interferometer are
located in two contiguous sections of the focal plane of the
SAGACE telescope. This enables angular differentiation of the sky
brightness, as detailed in the top-right inset} \label{aba:fig4}
\end{figure}

SAGACE is mainly a mm/sub-mm spectroscopic mission, taking
advantage of the high broad-band sensitivity of cryogenic
bolometeric detectors. To achieve a wide spectral range (100-760
GHz) with imaging capabilities and components with proven
readiness for space use, a Fourier Transform Spectrometer (FTS)
has been selected. In particular, we are considering as the
baseline a Martin-Puplett architecture \cite{Mart70}, because it
allows a clean differentiation of the spectra coming from two
independent inputs. In our case the two inputs cover two
contiguous areas of the focal plane, so that the measured spectrum
is the difference between the spectrum of the target source and
the spectrum of an offset reference field. This is sketched in
Fig.~\ref{aba:fig4}.

With this differential configuration, very small signals (like the
SZE distortion of the CMB in the direction of a cluster) can be
extracted from an overwhelming common mode background, generated
by the CMB itself, plus the emission of the warm telescope
(radiatively cooled to about 80~K), plus most of the spillover
from the Earth. The FTS has the important advantage over
dispersion spectrometers of being an imaging instrument. 2D
detector arrays can be accommodated in the focal plane, boosting
the mapping speed.

At these wavelengths, the achievable angular resolution is limited
by diffraction at the entrance aperture of the telescope. The
largest size of a primary mirror which can fit the Soyuz bay is
about 3.0~m; the rest of the available diameter is occupied by the
shields necessary to limit the sidelobes of the telescope, a
critical issue at these wavelengths. Budget constraints do not
allow us to consider deployable mirror solutions. We have
included in our baseline design a large (12~m diameter) deployable
Earth shield.

In order to limit the radiative background (and the corresponding
fluctuations) on the detectors, and to match the detector size to
the diffraction-limited angular resolution (which improves with
increasing frequency), we have divided our spectral range into four
bands. These bands have been optimized for performance in the study of
the SZE, taking into account the background emission from the warm
telescope and from the interstellar medium in our Galaxy.

The bands resulting from this optimization are $B_1 = 100-200$ GHz,
$B_2 = 200-300$ GHz, $B_3 = 300-450$ GHz, and $B_4 = 720-760$ GHz.
With a 3-m diameter dish the corresponding angular resolutions are
4.5, 2.25, 1.5, and 0.75~arcmin FWHM, respectively. A
$15' \times 15'$ field of view (resulting in two contiguous FOVs
in the sky, see Fig.~\ref{aba:fig4}) is filled with diffraction
limited detectors: the resulting number of detectors per band is
9, 36, 81, and 324. The photon background is dominated by the warm
telescope (at $\sim 80$~K), and is of order 0.8, 0.3, 0.2, and
0.1~pW in the four bands. The achievable photon-noise limited NEP
of each of the detectors is thus $(1.3, \ 1.0, \ 1.2, \ {\rm and} \
0.8) \times 10^{-17} \ \rm W/\sqrt{Hz}$.

The spectral resolution of the instrument depends on the maximum
delay introduced between the two beams of the interferometer. We
have selected a double pendulum configuration \cite{Burk83}, which
has been used several times in space missions, due to its
simplicity and reliability of the movement. This is very important
in a cryogenic implementation as ours. Our moving mirrors are
moved by tilting their supporting frame around a flexural pivot,
completely avoiding bearings. Resonance oscillation of the frame
requires little energy for motion control, thus maximising
cryogen lifetime. A double interferometer configuration, similar
in spirit to the one described in \cite{Carl81} has been designed,
so that all the power coming from the sky is processed by the
instrument. The maximum optical path difference (OPD) introduced
by the motion of the mirrors is 180 mm, resulting in 1-GHz
resolution over the full 100-760~GHz frequency range. A low-resolution
mode with 9~mm OPD and 20~GHz resolution is also available.

\begin{figure}
\center \psfig{file=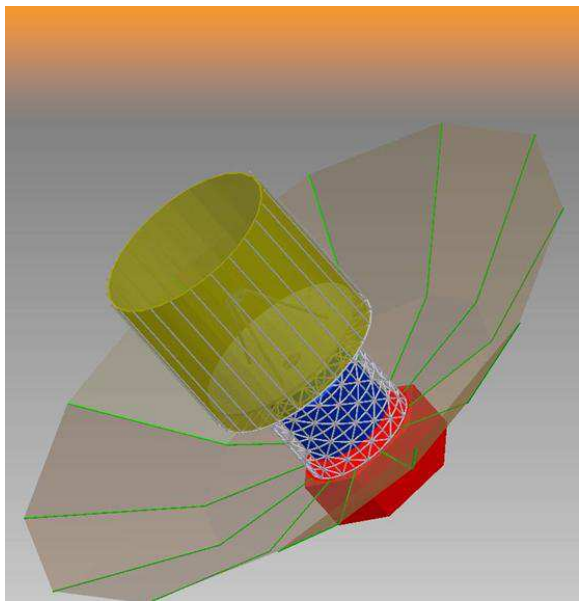,width=3in} \caption{Artist's
impression of the SAGACE satellite. The 3-m telescope is visible
through the inner shield, feeding the large cryostat (blue)
cooling the spectrometer and the detector arrays. The service
module is shown in red. The large deployable Sun/Earth shield is
12~m in diameter.} \label{aba:fig5}
\end{figure}

The detectors are fed by a fast Ritchey-Chr\'etien telescope, with
a 3-m diameter primary mirror. The entrance pupil is limited to
2.8~m in diameter by a cold Lyot stop. The secondary mirror
diameter is 45~cm, and the equivalent focal length of the
telescope is 9.2~m, while the distance between the subreflector
and the focal plane is only 1.47~m. The telescope is surrounded by
a fixed inner shield, and by a large (12~m diameter) deployable
Sun/Earth shield. The survey strategy is optimized to keep the
Sun, the Moon and the Earth at more than $90^\circ$ from the
telescope axis during observations. In this way, stray radiation
from these sources has to undergo two diffractions before hitting
the edge of the primary mirror. An artist's impression of the
SAGACE satellite is shown in Fig.~\ref{aba:fig5}.

For our telescope and spectrometer system, photon noise limited
bolometers in our bands reach NEFDs of 70, 52, 63, and 45~$\rm
mJy\sqrt{s}$ for each 20-GHz spectral bin in low-resolution mode,
and 1.4, 1.0, 1.3, and 0.9~$\rm Jy \sqrt{s}$ for each 1-GHz
spectral bin in high-resolution mode. The resulting performance in
terms of the science goals is reported in Section \ref{aba:perf}.

\section{Mission}

The SAGACE satellite needs to be three-axis stabilized, with a
pointing accuracy of 2~arcmin and a pointing knowledge of 15~arcsec,
with a stability of 1~arcsec/s and an agility of $90^\circ$ in
15~minutes.

The orbit of the SAGACE mission results from a cost/performance
trade-off. We want to avoid ground spillover as much as
possible. The best solution in this respect would be an L$_2$ orbit
similar to the Planck and Herschel orbits. We have, however,
investigated cheaper solutions, including Earth Sun-synchronous
orbits, medium circular orbits and elliptical orbits. The orbits
have been compared assuming that

\begin{itemlist}[(d)]
\item the telescope axis has to point at more than $135^\circ$ from the
 Sun and more than $90^\circ$ from the Earth's surface throughout
 the mission;
\item we want to reduce the spurious ground-diffraction signals
 below 1~$\rm \mu K$ during observations;
\item we want a duty-cycle (observation time/total time) larger than 50\%;
\item we do not want to spend a significant fraction of time
 in the inner radiation belt;
\item we do not want to have a propulsion system on board; and
\item we want to use only reaction wheels for the attitude control
 system and magnetic torquers as momentum-damping elements.
\end{itemlist}

The best trade-off we have found satisfying all these requirements
is a 6-hour low elliptical orbit, with 2364~km perigee and
18330~km apogee. The top three hours of the orbit are used for
science observation. During the other half-orbit the telescope
will point in the ``safe'' direction orthogonal to the orbit plane
and the magnetic torquers will perform their momentum damping
activity, allowing the speed of the flywheels to be maintained far
from saturation. The inclination of the orbit is $63.4^\circ$
(Molniya) to avoid the precession of the orbital plane. With a
southern-hemisphere apogee, we can use just one ground station
(the ASI-operated Malindi equatorial station) to control the
satellite and download the data. On average this gives a contact
time of 3.6~hours per day, split in four passes over the station,
at satellite altitudes between 4000 and 13000~km. For the low
resolution mode this is more than enough to download all the data
gathered by the instrument during the 12 hours per day of
observation. High resolution observations will alternate with
low-resolution observations to allow the download of high
resolution data stored on board.

The main disadvantage of this orbit is that it passes through the
inner radiation belt, while the apogee is located inside the outer
radiation belt. To mitigate the first problem, we implement
rad-hard technologies for the satellite and we switch several
subsystems off during the inner belt crossing. With respect to the
second problem, we have carried out detailed simulations of the
effect of cosmic rays on the bolometers in the outer radiation
belt. The simulation used GEANT-4 to compute the showers and
assumed a slab model where each bolometer is sandwitched between
two 3-mm layers of copper and two 10-mm layers of aluminium. To
simulate the radiation environment along the orbit we have
considered galactic cosmic rays (0.1-10~GeV), trapped electrons
(0.5-6.5~MeV) and trapped protons (20-50~MeV). We find that the
main contribution comes from galactic cosmic rays in the GeV
range, resulting in a rate of about 1 $\rm cm^{-2} \, s^{-1}$. The
small cross-section of spider-web bolometers then results in an
acceptable rate of glitches in the data.

The total dose absorbed by electronic components inside a 10mm
thickness equivalent aluminum box has been simulated using ESA's
SPENVIS SHIELDOLSE2 code, and is of the order of 10~krad in the
two-year mission.

\section{Observation Plan and Expected Performance}\label{aba:perf}

The SAGACE mission has been optimized for efficient conduct of the
investigations described in Section \ref{aba:sec1} in its two-year
lifetime, which is set by the cryogenic hold time. In particular,
18~months of the mission will be devoted to a low-resolution
($R=20$ at 300~GHz) survey of the sky, with 6~months on well-known
clusters and 12~months surveying blank sky regions, to discover
new clusters and to produce an important AGN catalogue, and a
survey of starforming galaxies over a broad redshift range with an
unprecedented combination of depth and area.

\begin{figure}
\psfig{file=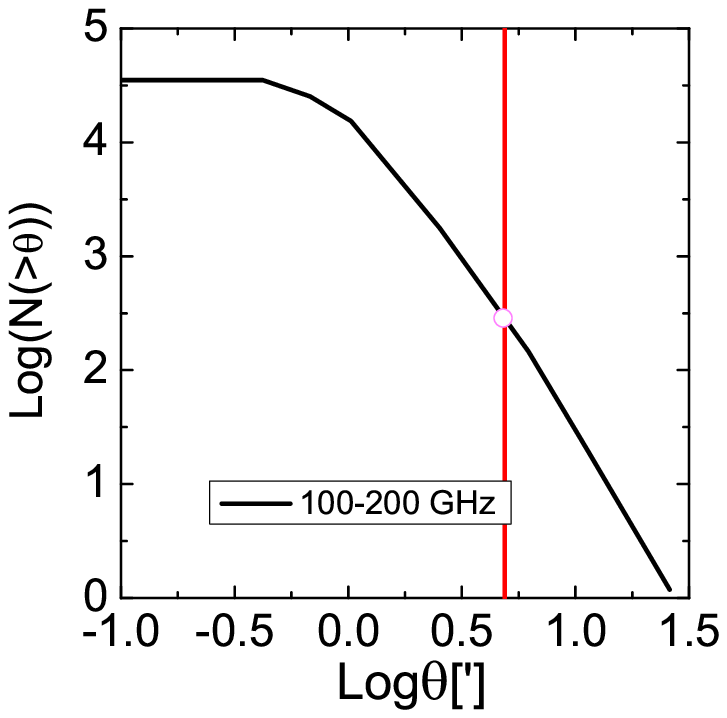,width=1.6in}
\psfig{file=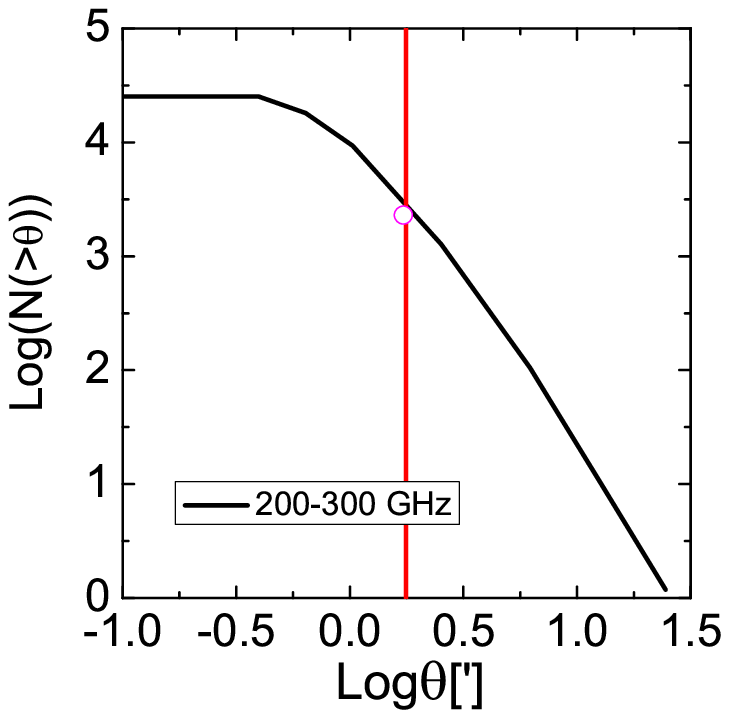,width=1.6in}
\psfig{file=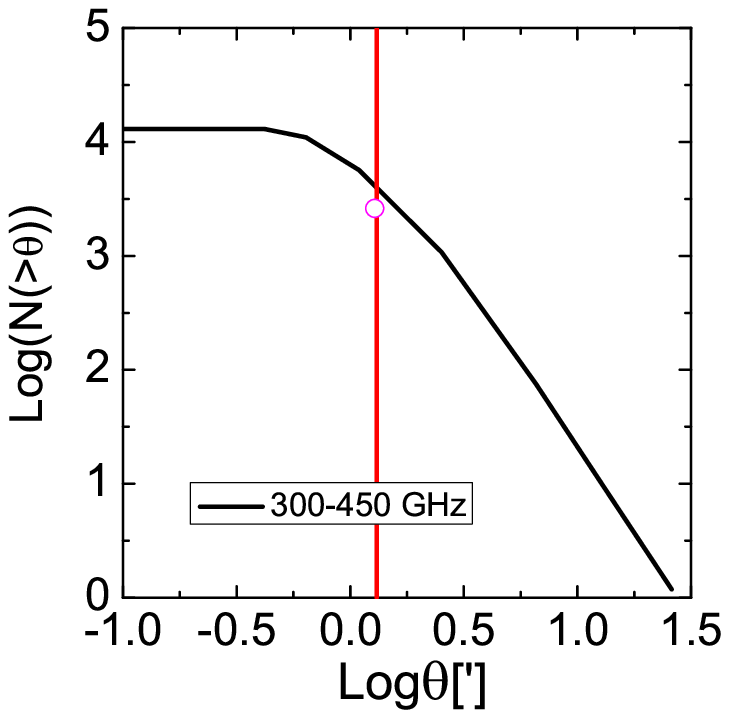,width=1.6in} \caption{All-sky cluster
counts vs.~intrinsic angular FWHM in the SAGACE bands 100-200 GHz
(left), 200-300 GHz (middle), and 300-450 GHz, (right). A limiting
flux density (per resolution element) of 20 mJy has been assumed.
The angular resolution of a telescope with a 3-m primary mirror is
marked by the vertical lines. Magenta dots show the expected
cluster counts taking into account the confusion from radio and
sub-mm unresolved sources.} \label{aba:fig6}
\end{figure}

The SAGACE instrument configuration will allow us to study
thousands of clusters in detail (see Fig.~\ref{aba:fig6}) with
full spectroscopic coverage in 100-450~GHz.

In Table~\ref{aba:tbl1} we report sample results from a detailed
Monte Carlo simulation of the recovery of parameters from SZE
observations of two clusters: a large nearby cluster (A1656, with
$z = 0.0230$, $\theta_c = 10.5^\prime$) and a small, distant
cluster (A0383, with $z = 0.187$, $\theta_c= 0.39^\prime$). We
have compared a ground-based three-band (95, 150, 225~GHz)
photometric measurement with characteristics as in\cite{Stan09},
the Planck survey using six bands for SZ detection and the other
three for foreground removal, and the SAGACE survey. The higher
resolution and longer integration time (with respect to Planck)
and the wide continuous spectral coverage of SAGACE lead to far
superior results. Not only are the errors greatly reduced (by a
factor 6-7 with respect to Planck for small clusters), but also
distributions of the recovered parameters are much less skewed,
implying a lower bias in their estimates.

\begin{table}
\tbl{Parameters Estimation Comparison}
{\begin{tabular}{@{}cccccc@{}}

\toprule Cluster & Parameter & input value & SPT & Planck & SAGACE
\\

\toprule

A1656 & $v_p ({\rm km/s})$ &  0 &  210$\pm$450 & 37$\pm$79 &
-31$\pm$32
  \\
  A1656 & $\tau$  & 0.00859 & 0.009$\pm$0.001 &  0.0088 $\pm$0.0002
   & 0.0085 $\pm$0.0002 \\
  \colrule
  A0383 & $v_p ({\rm km/s})$ &  0 & 10$\pm$530 & -410$\pm$910 &
   -20$\pm$140 \\
  A0383 & $\tau$ & 0.01924  & 0.025$\pm$0.007 &  0.0127$\pm$0.0077 &
   0.0186$\pm$0.0011 \\

 \botrule
\end{tabular}
} \label{aba:tbl1}
\end{table}

\begin{figure}
\center \psfig{file=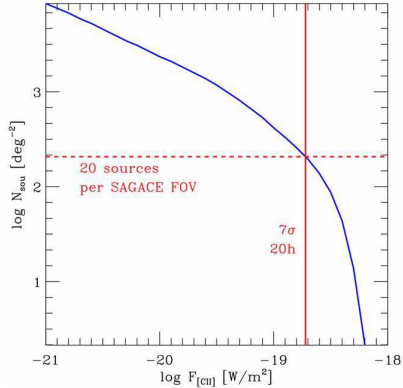,width=3in} \caption{Cumulative
counts of galaxies with [CII] line flux above a given level in the
720-760~GHz band (the highest frequency band of SAGACE) predicted
by the model in\cite{Lapi06}. The red line shows the expected
7$\sigma$ detection limit at high resolution ($R=740$) with an
integration time of 20~hours. With such an integration we can
sample the knee of the counts, i.e., maximize the number of
detected sources and sample those galaxies dominating the line
emission at z $\sim$ 1.5.} \label{aba:fig7}
\end{figure}

The remaining six months of the mission will be devoted to a
high-resolution ($R=700$ at 700~GHz) survey whose main output will
be a catalog of galaxies in the redshift desert detected in the
[CII] line. In Fig.~\ref{aba:fig7} we show how many galaxies it is
possible to detect in this way.

\section{Conclusions}

We have studied the implementation of a spectroscopic survey of
the mm/sub-mm sky on a small (cost-wise) space mission. Taking
advantage of the differential design of the spectrometer, we have
shown that a sensitive mission can use an elliptical Earth orbit,
reducing the cost of the launcher and of the attitude control
system. The two blind surveys to be performed by SAGACE, and the
extensive plan of pointed observations of galaxy clusters, will
provide a unique database for cosmological and astrophysical
studies of cosmic structures. Such a database will have a number
of applications for cosmology, high-energy astrophysics and
astro-particle physics.

\section*{Acknowledgements}

The phase-A study of SAGACE has been supported by the Italian
Space Agency (ASI) and by the Balzan Foundation.

\end{document}